\documentclass[singlecolumn,amsmath,amssymb]{revtex4}
\usepackage{graphicx}
\usepackage{bm, color}

\begin{document}
\title{Possible link between the distribution of atomic spectral lines
\\ and the radiation-matter-equilibrium in the early universe}

\author{Tim Richardt$^1$, Matthias Heinrich$^1$, Markus Gr\"afe$^2$, and Alexander Szameit$^{1,\ast}$}

\affiliation{
$^1$Institut f\"ur Physik, Universit\"at Rostock, Albert-Einstein-Stra{\ss}e 23, 18059 Rostock, Germany \\
$^2$Fraunhofer Institute for Applied Optics and Precision Engineering, Albert-Einstein-Str. 7, Jena, Germany\\
$^{\ast}$alexander.szameit@uni-rostock.de
}

\begin{abstract}
	\noindent\textbf{Abstract:} We report on the uncanny resemblance of the
	global distribution of all experimentally known atomic spectral lines to the
	Planckian spectral distribution associated with black body radiation at a
	temperature of $T\approx9000\,\mathrm{K}$. This value coincides with the
	critical temperature of equilibrium between the respective densities of
	radiation and matter in the early universe.
\end{abstract}

\maketitle

\noindent\textbf{The characteristic spectral lines are commonly understood to be
determined by the inner structure of an atom. As
such, they constitute a direct outcome of quantum mechanics. Apart from
higher-order corrections such as the Lamb shift, relativistic considerations
therefore typically do not come up when the distribution of discrete
electronic energy levels is discussed. In our work, we report on the observation
that the global distribution of all experimentally known atomic spectral lines
can be very accurately approximated by the Planckian function associated with
a temperature of $T\approx 9,000\,\mathrm{K}$. As it happens, this value
coincides with the critical temperature at which the respective densities of radiation density matter
were in equilibrium in the early universe
\cite{Gamov1948a, Gamov1948b, Alpher1948a, Alpher1948b}.
Since we cannot present any explanation for our findings yet,
this observation remains, for the moment, a striking conundrum. However, as so
aptly stated by the late Isaac Asimov, \emph{``the most exciting phrase to hear in science, the one that heralds new discoveries, is not `Eureka!' (I found
it!) but `That's funny\ldots '\,''}. In this spirit, we hope that an
interested reader may possibly find a more satisfying explanation beyond the
default assumption of a spurious correlation.}

\begin{figure}[h!]
\includegraphics[width=0.5\linewidth]{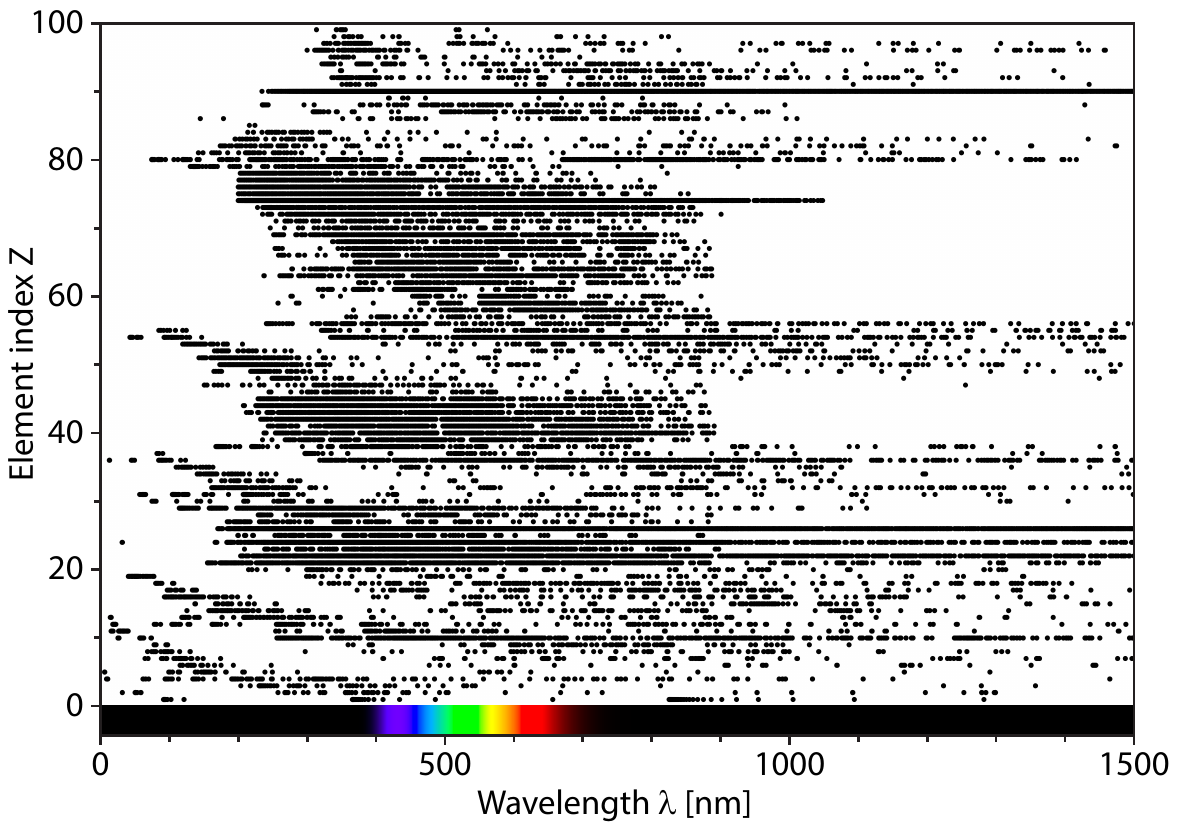}
\caption{Overview of the known and experimentally verified
spectral lines of neutral atoms up to $Z=100$ as listed in the NIST online catalogue \cite{NIST}. \label{fig:1}}
\end{figure}

The first spectral lines were measured by William H. Wollaston and
Joseph von Fraunhofer in 1802 and 1814, respectively
\cite{Wollaston1802,Fraunhofer1814}. Initially, the presence of these features in the spectrum of the sun was a puzzle to the scientific community. Almost a whole century passed until Niels Bohr put forward his famous
atomic model \cite{Bohr1913} as an explanation, which was later substantiated by
Louis de Broglie \cite{deBroglie1924}. Today, using the Schr\"odinger equation
\cite{Schrodinger1926}, atomic spectral lines (ASL) can be predicted with
extremely high accuracy for numerous elements and their isotopes.
Currently, more than 250,000 ASL are known and have been experimentally
verified. The National Institute of Standards and Technology (NIST) manages an
online catalogue wherein each of these lines is documented \cite{NIST} and
-- barring complications such as government shutdowns -- freely accessible.
A global overview of this impressive data set is shown in Figure \ref{fig:1}.

\begin{figure}
\includegraphics[width=0.5\linewidth]{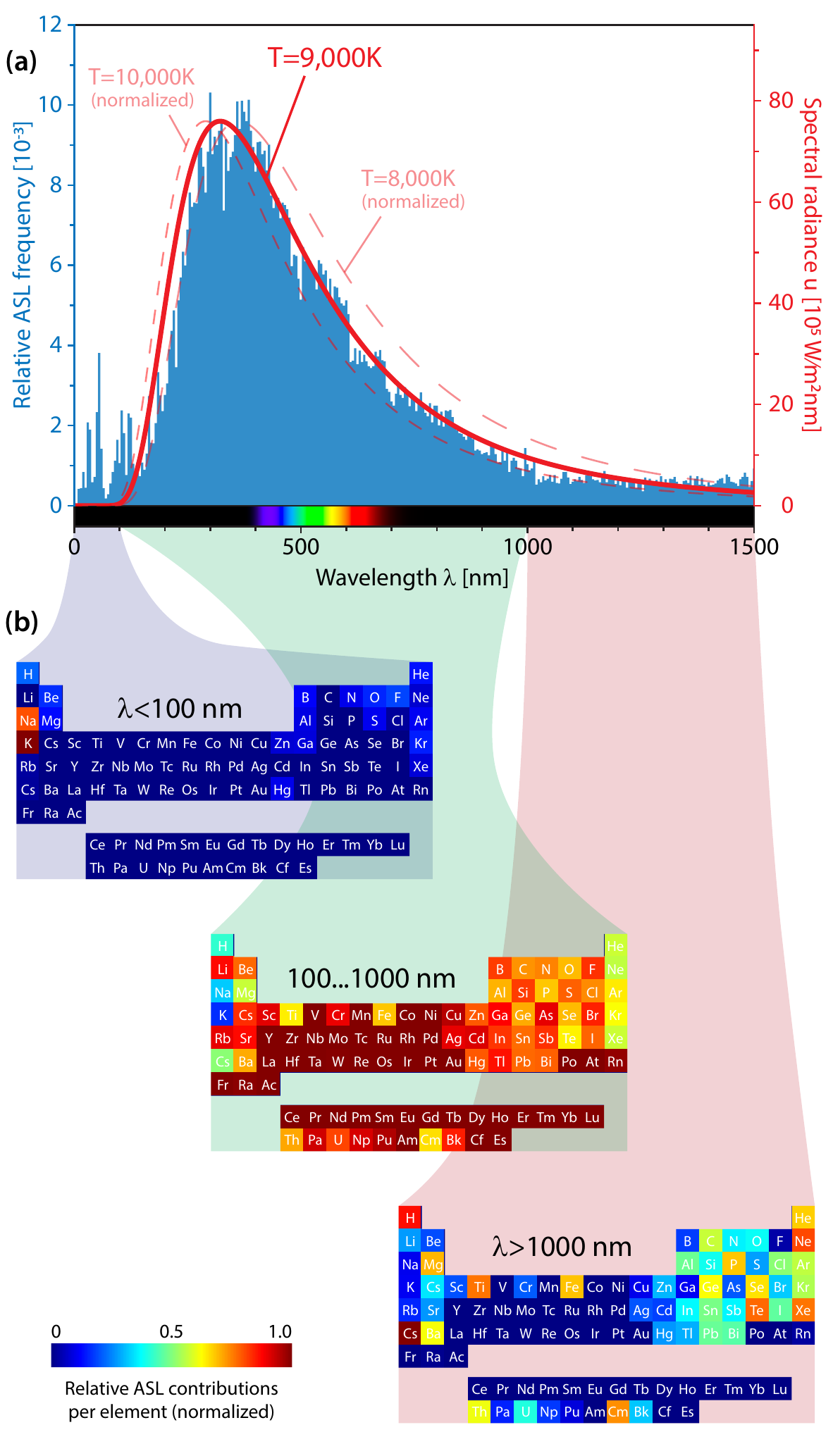}
\caption{(a) Histogram of the relative frequency of all experimentally verified atomic spectral lines
with a bin size of $5\,\textrm{nm}$ (blue bars). Its envelope closely traces the shape of the Planck distribution of the spectral radiance associated with a black body at the temperature $T=9,000\,\textrm{K}$ (solid red line) with its peak position of $\lambda=320\,\textrm{nm}$. The Planck distributions for $T=8,000\,\textrm{K}$ and $T=10,000\,\textrm{K}$, each normalized to the same peak height, are shown as dashed red lines and highlight the best match at $9,000\,\textrm{K}$.\\
(b) Color-coded relative contributions to the number of atomic spectral lines from each element in the extreme ultraviolet ($\lambda<100\,\mathrm{nm}$), between $100\,\mathrm{nm}$ and $1000\,\mathrm{nm}$, as well as the long infrared tail ($\lambda>1000\,\mathrm{nm}$).
While outliers at short wavelengths predominantly stem from only two elements ($_{19}\mathrm{K}$ Potassium and $_{11}\mathrm{Na}$ Sodium), the long-wavelength tail of the histogram mainly comprises lines from main-group elements. Notably, elements from across the entire periodic table contribute evenly to the Planck-like shape of the histogram, which therefore does not seem to be the result of inadvertent selection bias. \label{fig:2}}
\end{figure}

At first glance, the ASL of the different elements appear to be essentially
randomly distributed. For some element numbers, a vast
number of lines are known (e.g. $13,387$ for $_{90}\mathrm{Th}$ Thorium),
for others only a few (two for $_{85}\mathrm{At}$ Astatine). Clearly,
natural abundance and stability are only some of the factors determining
the number of known lines for a given element, and a certain selection
bias informed by the degree of technological relevance is likely to be at play.
However, aside from these outliers, the overall size of the database
all but ensures a fair sampling across the periodic table. Whereas the specific wavelengths, at which ASL can be found, vary greatly
across the board, it is evident that their overwhelming majority falls in the band between
$100\,\textrm{nm}\lesssim\lambda\lesssim 1000\,\textrm{nm}$.

This becomes even more obvious from a spectral histogram of the
relative frequency (in the statistical sense) of the ASL, which is shown in Fig. \ref{fig:2}(a).
Here, the data presented in Fig. \ref{fig:1} is aggregated with a bin size of $5\,\textrm{nm}$.
Only a small fraction of ASL occurs at very small wavelengths in or below the extreme ultraviolet regime ($\lambda\lesssim100\,\mathrm{nm}$).
The distribution peaks around $\lambda\approx320\,\textrm{nm}$, whereas for higher wavelengths,
the relative frequency decreases once more. Altogether, the distribution of the
spectral lines' relative frequency bears a striking similarity to the Planck
distribution $u(\lambda)$ that describes the radiation energy density per wavelength unit
emitted by a black body \cite{Planck1901}
\begin{equation}\label{eq:planck}
	u(\lambda) = \frac{2\pi hc^2}{\lambda^5} \frac{1}{e^{\frac{hc}{\lambda k_B T}}-1} \,,
\end{equation}
where $c$, $h$ and $k_B$ are the vacuum speed of light, Planck's
constant and the Boltzmann constant, respectively.
As is well known, the peak position of this distribution is measure of the black
body temperature $T$ \cite{Wien1893}. A best match to our histogram
of known ASL is obtained for a temperature of $T\approx9,000\,\textrm{K}$; the
corresponding graph is plotted as red solid line in Fig. \ref{fig:2}(a).
While the match is particularly evident in the visible regime, small
deviations occur at small and very large wavelengths. Nevertheless, the
histogram of known ASL appears to be extraordinarily well described by
radiation emitted by a black body of a temperature of $9,000\,\textrm{K}$.

At its face, the similarity between the ASL histogram and the Planck distribution is entirely unexpected and may appear entirely coincidental, since the continuous spectrum of thermal black-body radiation is the antithesis of the discrete energy quanta associated with atomic spectra. Moreover, apart from their common root within classical quantum theory, the ASL of different elements share no obvious connection to one another that would explain their collective overall distribution. Nevertheless, in the proverbial bird's eye view enabled by the NIST database, their totality coalesces into the observed compact, continuous and distinctly shaped envelope. As shown in Figure \ref{fig:2}(b), the entire periodic table contributes evenly to the bulk of the histogram, whereas outliers in the extreme UV predominantly stem from only two elements ($_{19}\mathrm{K}$ Potassium and $_{11}\mathrm{Na}$ Sodium).
Inadvertent selection effects of the aforementioned discrepancies in the number of known ASL for individual elements therefore do not lend themselves as convenient explanation for the observed similarity. As we will outline in the following, the seemingly arbitrary ``temperature'' of $9000\,\mathrm{K}$ --  notably the only degree of freedom in Eq.\eqref{eq:planck} -- may suggest a startling connection to cosmology.

Soon after Albert Einstein presented the fundamental concepts for general
relativity in 1916 \cite{Einstein1916}, Alexander Friedmann derived his famous
equation governing the dynamics of a homogeneous, isotropic universe \cite{Friedmann1922}:
\begin{equation}\label{eq:Friedmann}
	\left ( \frac{\dot{a}}{a}\right )^2 = \frac{8\pi G}{3} \rho(t) - \frac{kc}{a^2}
+ \frac{1}{3} \Lambda \; ,
\end{equation}
where $G$ denotes the gravitational constant, $\Lambda$ represents the cosmological constant, $k$ is a constant relating to the curvature, and $a$ is the scale factor of the universe, essentially describing its size. From this equation, he was able to derive the
energy-continuity equation
\begin{equation}
	\dot{\rho} = -3 \left ( \rho + p \right ) \frac{\dot{a}}{a} \; ,
\end{equation}
which connects pressure $p$ and energy density $\rho$. On these grounds, he
considered two possible extreme situations: The first is a universe dominated by
matter, such that $p=0$ and, consequently,
\begin{equation}
	\rho_{\textrm{mat}}\varpropto a^{-3} \; .
\end{equation}
The second is a universe dominated by radiation, such that $p=\rho/3$ and,
consequently,
\begin{equation}
	\rho_{\textrm{rad}}\varpropto a^{-4}\; .
\end{equation}
Although today our universe is evidently matter-dominated, due to the different
scaling of the two terms there must have been a time when $\rho_{\textrm{mat}} =
\rho_{\textrm{rad}}$.
Using Eq. \ref{eq:Friedmann}, one can show that this transition occurred at a quite young age of the universe,
approximately $50,000$ years after the big bang \cite{Carroll2007}. As
it turns out, this phase transition took place when the temperature of the universe was approximately
$T\approx9,000\,\textrm{K}$ \cite{Gamov1948a, Gamov1948b, Alpher1948a,
Alpher1948b}.

Setting aside the distinct possibility of a spurious correlation for the moment,
a possible interpretation would be that this phase transition
in the early universe may have played a previously unknown role in the formation of matter.
However, the fundamental characteristics of elementary particles that determine the electromagnetic properties of the elements were already in place at this point. Avoiding the pitfalls of the \textit{post hoc} fallacy \cite{logicalfallacy}, a yet-to-be-discovered, deeper causal linkage may have left its fingerprints both in the distribution of atomic spectral lines across the elements of
the periodic table, as well as the energy-matter balance of the universe as a whole. While we currently cannot offer a physical explanation to our entirely heuristic findings, it is our hope
that they may inspire future work that could eventually lead to the resolution of this fascinating puzzle.

\section*{Methods}

\noindent The NIST Atomic Spectra Database provides a comprehensive list of all
experimentally verified spectral lines and line energies \cite{NIST}.
We retrieved this data with a self-written program
\cite{TimsProgramm} and evaluated the number
of ASL in wavelength intervals of $5\,\textrm{nm}$. Notably, as long as dramatic
oversampling is avoided, the specific window size neither has a systematic impact on
the shape of the distribution, nor the on the temperature associated with the best-match black body spectrum.

The relative ASL contributions depicted in Figure \ref{fig:2}(b) were determined by calculating the fraction of known spectral lines of a given element that fall in the spectral range under consideration. For example, of the 214 lines listed for $_1\mathrm{H}$ Hydrogen, 37 fall below $100\,\mathrm{nm}$ ($17.3\%$), 91 in the range between $100...1000\,\mathrm{nm}$ ($42.5\%$), and the remaining 86 lines ($40.2\%$) above $1000\,\mathrm{nm}$. In calculating these relative contributions, we were able to meaningfully compare the impact of elements with hugely disparate numbers of known lines.

\begin{figure}[h!]
\includegraphics[width=0.5\linewidth]{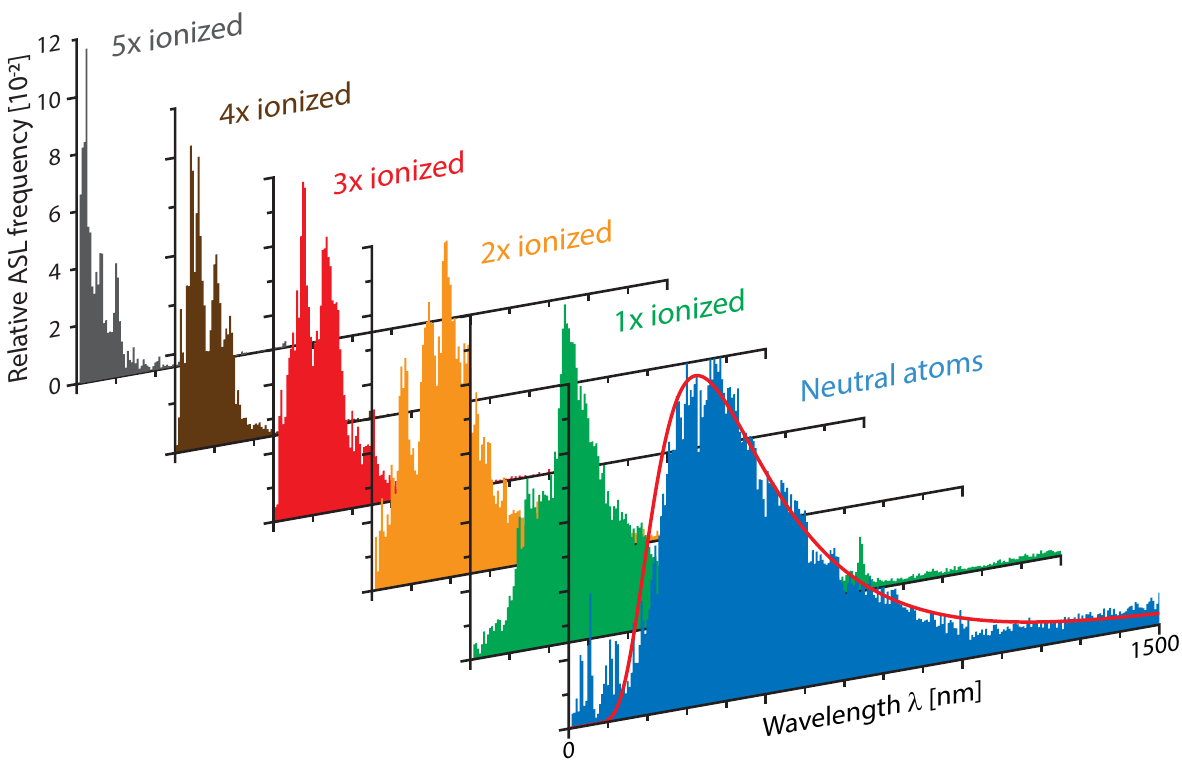}
\caption{Histograms for ionic spectral lines. Each panel corresponds to an ionization step. The effectively larger nuclear charge of ions shifts their spectra towards shorter wavelengths. At the same time, the increasing sparsity of the data set yields more jagged histograms. To avoid these systematic and stochastic distortions, we excluded ionic spectral lines from our statistical analysis. \label{fig:3}}
\end{figure}

In our statistical analysis, we focused on the spectral lines of neutral atoms, since each successive step of ionization necessarily excludes an element from the statistics and therefore over-emphasize the impact of higher elements. In a similar vein, each remaining electron in a given ion experiences a systematically deeper potential well, as the nuclear charge is less shielded, resulting in the histogram of spectral lines being shifted towards lower wavelengths (see Figure \ref{fig:3}). Finally, the available spectral data becomes increasingly sparse with higher ionization numbers, effectively omitting large swaths of the ion triangle. In light of these considerations, we decided against the inclusion of ionic spectral lines.

\section{Author contributions}
\noindent T.R. conceived the idea, retrieved and processed the data. T.R., M.H., M.G. and A.S. discussed and interpreted the results. M.H. and M.G. prepared the figures. A.S. and M.H. wrote the manuscript.

\section{Competing financial interests}

\noindent Competing interests, financial or otherwise, do not exist.

\end{document}